\documentclass[%
 reprint,
 amsmath,amssymb,
 aps,
]{revtex4-2}

\usepackage{graphicx}
\usepackage{dcolumn}
\usepackage{bm}
\usepackage[utf8]{inputenc}%
\usepackage{amssymb}
\usepackage{siunitx}
\usepackage{amsmath}
\usepackage{float}
\usepackage{hyperref}
\hypersetup{hidelinks,colorlinks=true,linkcolor=blue,citecolor=blue,urlcolor=black}
\usepackage{graphicx}
\usepackage{dcolumn}%

\begin{document}

\preprint{APS/123-QED}

\title{Tunable Magnetism and Valleys in VSiZ$_3$ monolayers}

\author{Xiaoyu Wang}
\author{Li Liang}%
\author{Huiqian Wang}
\author{Xiao Li}
\email{lixiao@njnu.edu.cn}
\affiliation{%
 Center for Quantum Transport and Thermal Energy Science, School of Physics and Technology,
Nanjing Normal University, Nanjing 210023, China}%

\begin{abstract}
Two-dimensional magnetism and valleys have recently emerged as two significant research areas, with intriguing properties and practical uses in advanced information technology. Considering the importance of these two areas and their couplings, controllable creations of both the magnetism and valley polarization are highly sought after. Based on first-principles calculations, we propose a new class of two-dimensional monolayers with a chemical formula of MAZ$_3$, which is viewed as a 2H-MZ$_2$ trilayer  passivated by the A-Z bilayer on its one side. Taking VSiN$_3$ as an example, the MAZ$_3$ monolayers are found to exhibit tunable magnetism and valleys. For the intrinsic VSiN$_3$ monolayer, it is a non-magnetic semiconductor, with multiple degenerate valleys and trigonal warping near \emph{K$_\pm$} points in the band structure. Besides, the bands have spin splittings owing to the spin-orbit coupling. Under a moderate carrier doping, the monolayer becomes a Stoner ferromagnet, which enhances the spin splittings of the valence band and generates valley splittings. Moreover, the Berry curvature is valley contrasting, leading to distinct valley-spin related anomalous Hall currents as the doping concentration increases. Our work opens up new way to modulate the spin splittings and valley splittings via electric means, and provides opportunities for exploring advanced spintronic and valleytronic devices.

\end{abstract}

\maketitle


\section{\label{sec:level1}INTRODUCTION}

 Two-dimensional magnetic materials have been a rapid developing field, exhibiting conceptual importance and enormous applications  \cite{CGT,CRI3,RECENT2DPROGRESS,MAGNETICMATERIALS,DEVICES}. Besides intrinsic magnetic materials \cite{CGT,CRI3}, the magnetism can be introduced into two-dimensional materials by external manipulations \cite{GASE,IN3SE3,BIGRAHNENE}. For example, the carrier doping serves as an effective, controllable method, which induces the magnetism in e.g. the GaSe monolayer  \cite{GASE}, the In$_2$Se$_3$ monolayer \cite{IN3SE3} and the graphene bilayer \cite{BIGRAHNENE}. On the other hand, valley degree of freedom has also received considerable attentions in recent years \cite{Valleytronicsin2Dmaterials,VALLEY}. In a two-dimensional valley material, inequivalent valleys present contrasting optical, magnetic and electronic properties \cite{MagneticMomentandTopologicalTransport,opt,circulardichroism,YAOWANG}. The valleys can thus be used as information carriers in information technologies. Moreover, the valley structure is closely related to the magnetism, and it is highly tunable by applied magnetic field and magnetic orders \cite{PNAS,15-6,15-7,15-8,15-9,15LIXIAO,ferrovalley,LIANGLI}. Given the significance of two-dimensional magnetism and valley degree of freedom, it is highly desirable to realize controllable modulations of both the magnetism and valleys in a two-dimensional material.

Recently, a new family of two-dimensional materials, MA$_2$Z$_4$ (M $=$ transition metal; A $=$ Si, Ge; Z $=$ N, P, As), has been predicted theoretically and then synthesized experimentally \cite{MOSI2N4,MA2Z4}. The MA$_2$Z$_4$ monolayer can be regarded as a MZ$_2$ trilayer in the 2H phase passivated by A-Z bilayers on its two sides. Subsequently, the Si-N passivated graphene has been investigated, where the graphene is sandwiched between two Si-N bilayers, or passivated by the Si-N bilayer on its one side \cite{SI-N}. A variety of intriguing physical properties, e.g. superconductivity and band topology, are manifested in the materials with adherent A-Z bilayers \cite{MOSI2N4,MA2Z4,SI-N,LIANGLI}. However, in contrast to the case of the graphene, the single-sided passivation of the 2H-MZ$_2$ trilayer with the A-Z bilayer has not been studied, and its stabilities and physical characteristics are worth further research. 


In this work, we focus on the 2H-MZ$_2$ trilayer passivated by the A-Z bilayer on its one side, which has a chemical formula of MAZ$_3$. Taken the VSiN$_3$ monolayer as an example, we study its structural and electronic properties by the first-principles calculations. Our calculations demonstrate the monolayer is dynamically and thermodynamically stable. The intrinsic VSiN$_3$ monolayer is a non-magnetic semiconductor, exhibiting degenerate valleys and trigonal warping effect around $K_\pm$ in its electronic band structure. The valley contrasting Berry curvature appears, enabling valley-related anomalous Hall transport. Besides, the spin-orbit coupling gives rise to spin splittings. More interestingly, the carrier doping creates the Stoner ferromagnetism in the monolayer, owing to large densities of states at band edges. The induced magnetism further leads to valley splittings and enlarges the spin splittings of the valence band. Therefore, the magnetism and valleys in VSiN$_3$ monolayer are highly tunable by electric means. These findings broaden the scopes of the researches of two-dimensional magnetic materials and valley materials, and pave an avenue for designing energy-efficient information storage and processing devices.


\section{METHODS}
Our first-principles calculations are performed within the framework of the density functional theory \cite{DFT1,DFT2}, as implemented in Vienna Ab initio Simulation Package \cite{VASP1,VASP2}. The projector augmented wave potentials \cite{PAW} and the Perdew-Burke-Ernzerhof exchange-correlation functionals \cite{PBE} are used in our calculations, with a plane-wave energy cutoff of 500 eV. A $\Gamma$-centered \textbf{\textit{k}} mesh of 15$\times$15$\times$1 is adopted to sample the Brillouin zone. Atomic structures are relaxed until the interatomic force on each atom is less than 0.001 eV/Å. The electronic iteration convergence criterion is set to $10^{-8}$ eV. A vacuum layer of more than 17 \AA \ is inserted to avoid the interaction between the VSiZ$_3$ monolayer and its periodical image. The $d$ orbitals of the V atom are treated using the GGA$+U$ approach with an effective Hubbard $U$ of 3 eV \cite{GGAU,V3}. The spin-orbit coupling is introduced into the electronic structure calculations, unless otherwise specified. The effect of the charge doping is simulated by changing the total number of electrons in the unit cell, along with a compensating jellium background of opposite charges that maintains the neutrality of the system. The phonon spectrum is calculated within the density functional perturbation theory using the PHONOPY code \cite{phonopy}. The ab initio molecular dynamics (AIMD) with Nosé-Hoover thermostat is applied to evaluate the thermal stability of the VSiZ$_3$ monolayer \cite{AIMD}, where a 5$\times$5$\times$1 supercell is used to reduce the constraint from the periodicity.

To quantify the spin splitting, $\Delta^{v/c,\textbf{\emph{k}}}_\text{spin}$, for the valence or conduction band at a given wave vector \textbf{\emph{k}} and the valley splitting, $\Delta^{v/c}_\text{val}$, for a given band in the electronic band structure, we define the relations
\begin{eqnarray}
{\Delta^{v/c,\textbf{\emph{k}}}_\text{spin}}={E^{v/c,\textbf{\emph{k}}}_{\uparrow}}-{E^{v/c,\textbf{\emph{k}}}_{\downarrow}} ,
\\
{\Delta^{v/c}_\text{val}}={E^{v/c,+1}_\text{val}}{-E^{v/c,-1}_\text{val}}.
\end{eqnarray}
Here, \emph{v} and \emph{c} represent the valence and conduction bands, respectively. $\uparrow$ and $\downarrow$ denote the spin-up and spin-down states, respectively. $E^{v/c,\textbf{\emph{k}}}_{\uparrow/\downarrow}$ in Eq. (1) is the energy of the corresponding electronic state at a certain \textbf{\emph{k}}. $E^{v/c,+1}_\text{val}$ and $E^{v/c,-1}_\text{val}$ in Eq. (2) are the energies of the valence/conduction band states at $X_+$/$Y_-$ and $X_-$/$Y_-$ valleys, respectively.

\section{RESULTS}
\subsection{\label{sec:level2}Atomic Structure and Stability of 
\\
the VSiN$_3$ Monolayer}

Figs. 1(a) and (b) show the atomic structure of the VSiN$_3$ monolayer. The monolayer has a two-dimensional hexagonal lattice, with the three-fold rotational symmetry. It consists of five atomic layers stacked as N-V-N-Si-N. The upper Si-N bilayer forms a buckled honeycomb structure. As for the lower N-V-N trilayer, each V atom has six neighboring N atoms that compose a triangular prism, which is similar to the atomic structure of the 2H-MoS$_2$ monolayer. The VSiN$_3$ monolayer is regarded as the passivation of the N-V-N trilayer by the Si-N bilayer on its one side. This is in contrast to widely studied VSi$_2$N$_4$ monolayer \cite{1vsi2n4,2vsi2n4}, where two Si-N bilayers are attached to two sides of the N-V-N trilayer. For the VSiN$_3$ monolayer, the in-plane lattice constant of its hexagonal lattice is computed to be 2.90 \AA, and the thickness of the monolayer is 4.48 \AA. We also performed the spin-polarized density functional theory calculations, and it is found that the monolayer is nonmagnetic. 

\begin{figure}[htbp]
\includegraphics[width=8.5 cm]{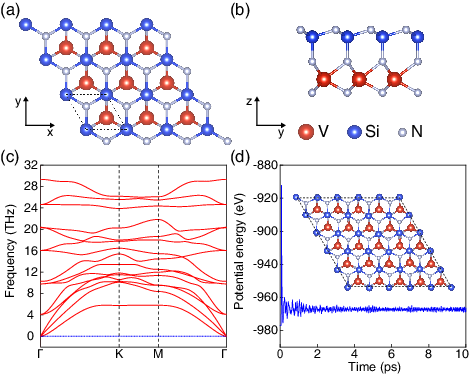}
\caption{\label{fig:epsart}Atomic structure and stability of the VSiN$_3$ monolayer.(a) The top view and (b) the side view of the monolayer. Red, blue, and gray balls stand for V, Si, and N atoms, respectively. (c) The phonon dispersion. (d) Energy variation at 600 K using ab initio molecular dynamics simulations. The inset shows the atomic structure after 10 ps.}
\end{figure}

We then investigate the structural stability of the VSiN$_3$ monolayer, including its dynamical and thermodynamical stabilities. The results of the phonon spectrum and AIMD simulations of the  VSiN$_3$ monolayer are shown in Figs. 1(c) and (d), respectively. According to the phonon spectrum, there is no obvious imaginary mode in the entire phonon Brillouin zone, confirming the dynamical stability of the monolayers. As for the AIMD simulation at 600 K for 10 ps, it is found that the total energy of the monolayer almost remains at a fixed value and the atomic structure is well kept, verifying the thermodynamical stability of the monolayer at a relatively high temperature.

\subsection{\label{sec:level2}Electronic Properties of the VSiN$_3$ Monolayer}

In the followings, we study electronic properties of the VSiN$_3$ monolayer. The electronic band structure of the monolayer are presented in Fig. 2(a), with the spin-orbit coupling considered. It is seen that there are spin splittings for electronic bands, which arises from the spin-orbit coupling. The spin splitting of the conduction band is considerable, with values of $\pm$61 meV at high symmetry \emph{K$_\pm$}, respectively. Similar to the case at \emph{K$_\pm$}, opposite spin splittings also occur at other pairs of opposite wave vectors, which are ensured by the presence of the time-reversal symmetry \cite{YAOWANG}. As for the valence band, the spin splitting is smaller, with magnitudes of less than 20 meV, which is magnified in the inset of Fig. 2(a). 

\begin{figure}[htbp]
\includegraphics[width=8.5 cm]{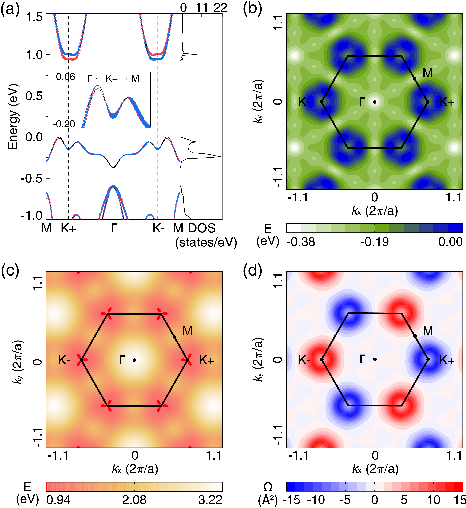}
\caption{\label{fig:epsart}Electronic properties of the VSiN$_3$ monolayer. (a) The band structure and corresponding density of states, where the spin-orbit coupling is taken into account. (b) The energy contour of the valence band. (c) The energy contour of the conduction band. (d) The Berry curvature in the momentum space. In (a), the spin-up and spin-down electronic states are represented by red and blue circles, respectively, with the size of the circle being proportional to the vertical spin component. The inset zooms in on the valence band around the \emph{K$_+$} point, exhibiting spin splittings arising from the spin-orbit coupling. The valence band maximum is set to zero energy. }
\end{figure}

We then focus on band edges states. The valence band maxima (VBM) and the conduction band minima (CBM) are located in the neighborhood of \emph{K$_\pm$} points. The corresponding wave vectors are in the \emph{K$_\pm$}-\emph{M} and \emph{K$_\pm$}-\emph{$\Gamma$} paths, and they are denoted as \emph{X$_\pm$} and \emph{Y$_\pm$}, respectively. Here, \emph{X$_+$} (\emph{Y$_+$})  is closer to \emph{K$_+$}, and \emph{X$_-$} (\emph{Y$_-$}) is closer to \emph{K$_-$}. The band edge states are degenerate at \emph{X$_+$} (\emph{Y$_+$}) valley and \emph{X$_-$} (\emph{Y$_-$}) valley, similiar to the \emph{K$_\pm$} points. Considering the energy difference between the VBM at \emph{X$_\pm$} valleys and CBM at \emph{Y$_\pm$} valleys, a global indirect band gap has a a magnitude of 0.94 eV. Furthermore, Figs. 2(b) and (c) provide the energy contours of the valence and conduction bands in the two-dimensional momentum space. It is seen there are three degenerate \emph{X$_\pm$} (\emph{Y$_\pm$}) valleys around \emph{K$_\pm$} owing to the presence of the three-fold rotation symmetry. As the energy moves away from the band edge, the energy contour first exhibits three pockets centered at \emph{X$_\pm$} (\emph{Y$_\pm$}), and then becomes distorted ring, exhibiting the trigonal warping effect and the Lifshitz transition of the topology of the isoenergetic contour \cite{Lifshitz}.

Given that there are \emph{X$_\pm$} and \emph{Y$_\pm$} valleys for the valence and conduction bands, respectively, we further explore possible valley contrasting properties in the VSiN$_3$ monolayer. Fig. 2(d) presents the Berry curvature in the two-dimensional momentum space, and the Berry curvature along the high symmetry paths is given in the Supporting Information (SI hereafter). The Berry curvature is calculated by considering the contributions from all occupied bands. Similar to the energy contour, the shape of the Berry curvature exhibit the three-fold rotational symmetry around \emph{K$_\pm$}. The \emph{X$_\pm$} valleys give sizable Berry curvatures that have the same magnitude but with opposite signs, owing to the spatial inversion symmetry breaking and the presence of the time-reversal symmetry \cite{Berry}. As a result, opposite anomalous Hall currents can be generated from \emph{X$_\pm$} valleys, under the action of the in-plane electric field, exhibiting the valley Hall effect \cite{YAOWANG}.

Moreover, we also show the density of states of the  VSiN$_3$ monolayer in Fig. 2(a), along with the electronic band structure. The density of states suddenly increases near the band edges, indicating the van Hove singularity, which results from the relatively flat dispersion of the band edges.

\subsection{\label{sec:level2}Tunable Magnetism and Electronic properties of Carrier-doped VSiN$_3$ Monolayer}

Given that the large density of states at the Fermi level may lead to the emergence of the magnetism and the carrier doping can shift the Fermi level, we further study the effect from the carrier doping on the magnetism of the VSiN$_3$ monolayer. Fig. 3(a) demonstrates the evolution of the magnetic moment per unit cell with the increasing concentration of the hole doping for the monolayer. Without the carrier doping, the intrinsic monolayer is nonmagnetic, as mentioned above. At a small amount of the hole doping, i.e. the doping concentration with the order of 0.01 hole per cell, the monolayer spontaneously develops a ferromagnetic ground state. As the doping concentration increases, the magnetic moment is almost linearly increased within the doping range considered in Fig. 3(a). The maximum doping concentration considered here, 0.25 hole per unit cell, gives a sizable magnetic moment of 0.25 $\mu_\text{B}$ per unit cell. The concentration corresponds to 3.42$\times10^{14}$ cm$^{-2}$, which can be readily realized experimentally \cite{EXPERIMENT1,EXPERIMENT2}. Therefore, the hole doping indeed induces the ferromagnetism in the VSiN$_3$ monolayer.

 Moreover, we also calculate the atomic contributions to the magnetic moment in a unit cell, which is provided in SI. It is found that for the VSiN$_3$ monolayer, the magnetic moments are mainly contributed by the V and N atoms, while the contribution from the Si atom is negligible. The magnetic moments contributed by V and N atoms are antiferromagnetically coupled. On the other hand, our calculations show the electron doping can also lead to the magnetism for the VSiN$_3$ monolayer, where the total magnetic moment also linearly increases with the doping concentration in the range below 0.25 electrons per unit cell (see SI).  

\begin{figure}[htbp]
\includegraphics[width=8.5 cm]{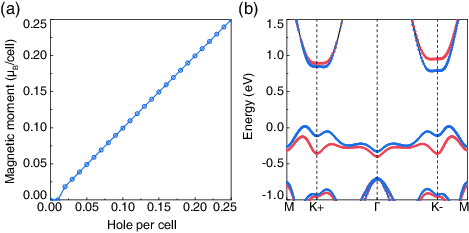}
\caption{\label{fig:epsart}The magnetic and electronic properties of the hole-doped VSiN$_3$ monolayer. (a) The evolution of the total magnetic moment in a unit cell with the increasing hole doping. (b) The electronic band structures at the doping concentration of 0.1 hole per unit cell.}
\end{figure}
To elucidate the effects of the carrier doping and induced magnetism on electronic properties of the VSiN$_3$ monolayer, we further plot the band structure with a doping concentration of 0.1 hole per cell in Fig. 3(b). The hole doping and associated magnetism introduce a magnetic exchange field, leading to relative shifts between the bands with opposite spins and the changes of spin splittings . The first and second valence bands become spin-down and spin-up ones along the entire high symmetry paths, respectively. Similarly, the first and second conduction bands are also spin-down and spin-up, respectively, in the neighborhood of $K_\pm$. That is, the spin splittings in the regions around $K_+$ and $K_-$ have the same sign, which is different from the spin splittings induced by only the spin-orbit coupling in Fig. 2(a). Besides, the spin splittings between the two valence bands become much larger than those in Fig. 2(a). The above findings indicate that the exchange interaction from the magnetism is stronger than the spin-orbit coupling in the doped monolayer. Furthermore, the simultaneous presences of the magnetism and spin-orbit coupling give rise to valley splittings, i.e. the valence (conduction) bands states at \emph{X$_+$} (\emph{Y$_+$}) and \emph{X$_-$} (\emph{Y$_-$}) valleys have different energies. According to the definition of the valley splitting in the Method section, the valley splittings of the valence band is computed to be 2 meV. In contrast, the conduction band has a much larger valley splitting, of which the magnitude is 56 meV. Besides, owing to the hole doping, the Fermi level goes across the first valence band around $X_\pm$ valleys. 

 The induced magnetism by the carrier doping can be qualitatively explained by the Stoner model. According to the Stoner model, when the Stoner criterion, $D(E_\text{F})J$\textgreater1, is satisfied, the Stoner ferromagnetism will occur. Here, $D(E_\text{F})$ is the density of states at the Fermi energy, $E_\text{F}$, in a non-magnetic state. $J$ is the strength of the exchange interaction, and it can be estimated by the ratio of the splitting between spin-up and spin-down electronic states near the Fermi level to the magnetic moment generating the spin splitting  \cite{IN3SE3,I}. Specific to the VSiN$_3$ monolayer with the doping concentration of 0.1 hole per cell, it is found that the induced magnetic moment is 0.1 $\mu_\text{B}$ in Fig. 3(a). As demonstrated in SI, the density of state, $D(E_\text{F})$, is 3.6 states/eV, and the spin splitting between the first and second valence bands is on the order of 0.1 eV. Therefore, $D(E_\text{F})J$ is larger than 1, which indicates the  physical origin of the Stoner ferromagnetism in the VSiN$_3$ monolayer.

\section{DISCUSSIONS}
As for the carrier-doped monolayers, we also calculate the momentum-resolved Berry curvature, which is still opposite at $X_\pm$ ($Y_\pm$) valleys for the valence (conduction) band. The valley Hall effect is well kept within the doping concentration considered. On the other hand, there are the sign changes of spin splittings at valleys with introducing the carrier doping. As mentioned above,  while the spin splittings have opposite signs at $X_\pm$ ($Y_\pm$) valleys for the intrinsic VSiN$_3$ monolayer, the signs of the spin splittings become the same, as the doping concentration increases. Because of the spin splittings above, the valley Hall effect is also associated with the spin and it undergoes a transition of the spin polarization. With increasing the doping concentration, the valley Hall currents are changed from the opposite spins to the single spin.    

Moreover, multiple $X_\pm$ ($Y_\pm$) valleys appear in the electronic band structure of the VSiN$_3$ monolayer, and their degeneracies are highly tunable. For the intrinsic monolayer, the valence (conduction) band states are degenerate at \emph{X$_\pm$} (\emph{Y$_\pm$}) valleys. The carrier doping and induced magnetism lead to the degeneracy lifting between \emph{X$_+$} ($Y_+$) and \emph{X$_-$} ($Y_-$) valleys. It is because the time-reversal symmetry is broken owing to the induced magnetism. On the other hand, the degeneracy between three \emph{X$_+$} (\emph{X$_-$}, \emph{Y$_+$} or \emph{Y$_-$}) valleys is kept even though the carrier doping is introduced, since the degeneracy is protected by the three-fold rotation symmetry that is immune to the carrier doping. The valley degeneracy can thus be further lifted by breaking the rotation symmetry via an applied uniaxial strain, which enables the multiple modulations of the valleys in the VSiN$_3$ monolayer.    

In addition, the VSiN$_3$ monolayer is one of material candidates of the MZ$_2$ trilayer passivated by the A-Z bilayer on its one side. A large number of MAZ$_3$ monolayers are expected, with M, A and Z adopting various elements. As another example of MAZ$_3$ monolayers, the VSiP$_3$ monolayer has also been investigated by first-principles calculations. The VSiP$_3$ monolayer has a similar atomic structure with the VSiN$_3$ monolayer, and its in-plane lattice constant is computed to be 3.46 \AA. The monolayer also exhibits dynamic and thermodynamic stabilities, which are presented in SI. Besides, it has no magnetic order as well. 
\begin{figure}[htbp]
\includegraphics[width=8.5 cm]{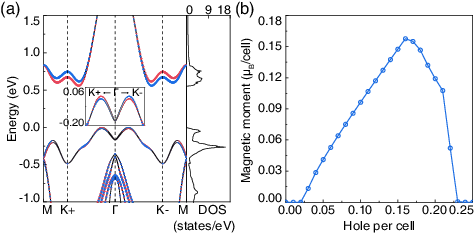}
\caption{\label{fig:epsart}Electronic and magnetic properties of the VSiP$_3$ monolayer, with considering the spin-orbit coupling. (a) The band structure and corresponding density of states. (b) The evolution of the total magnetic moment in a unit cell with the increasing hole doping. }
\end{figure}
Fig. 4(a) shows the electronic band structure and density of states of the intrinsic VSiP$_3$ monolayer. Similar to the VSiN$_3$ monolayer, the bands of the VSiP$_3$ monolayer are spin polarized. Different from VSiN$_3$, the VBM and the CBM are both located in the \emph{$\Gamma$}-\emph{K$_\pm$} paths, and they are near \emph{$\Gamma$} and \emph{K$_\pm$} points, respectively, leading to a global indirect band gap of 0.60 eV. As for the density of states, it is sizable at the band edges. We then study the effect from the carrier doping on the magnetism of the VSiP$_3$ monolayer. As shown in Fig. 4(b), the monolayer has an increased magnetic moment until the doping concentration of 0.16 hole per cell, with the largest magnetic moment of 0.16 $\mu_\text{B}$,  and then the magnetic moment sharply decreases to zero at the concentration of 0.23 hole per cell. Therefore, the monolayer also becomes a Stoner ferromagnet with a moderate carrier doping. The new characteristics found in the VSiZ$_3$ (Z $=$ N, P) monolayers are likely to have generalizations to more MAZ$_3$ monolayers. 
\section{CONCLUSION}
To summarize, taking the VSiN$_3$ monolayer as an example, we investigate the atomic and electronic structures of the MAZ$_3$ monolayers, by first-principles calculations. The intrinsic VSiN$_3$ monolayer is a non-magnetic semiconductor, with spin splittings induced by the spin-orbit coupling and the trigonal warping around $K_\pm$ in the electronic band structure. The Berry curvature around $K_\pm$ is valley contrasting, which can induce valley-related anomalous Hall transport. The large density of states at the band edges further enables the Stoner ferromagnetism via the moderate carrier doping. The induced magnetism gives rise to valley degeneracy splittings. Moreover, similar results have also been found in another MAZ$_3$ materials, such as the VSiP$_3$ monolayer. Our work provides a new family of two-dimensional materials with the carrier-doped induced Stoner ferromagnetism, and paves an avenue for the effective control of valleys via electric means.
\bigskip
\begin{acknowledgments}
We are grateful to Dr. Zhichao Zhou for valuable discussions. The work was supported by the National Natural Science Foundation of China (Grants No. 12374044 and No. 11904173) and the Jiangsu Specially-Appointed Professor Program.
\end{acknowledgments}


\nocite{*}

\end{document}